\documentclass[twocolumn,prb,showpacs,amsmath,amssymb]{revtex4}
\usepackage{graphicx}
\usepackage{color}
\usepackage{ulem}
\newcommand{\br}{\mathbf{r}}

\newcommand{\bk}{\mathbf{k}}

\newcommand{\bn}{\begin{equation}}
\newcommand{\ee}{\end{equation}}
\newcommand{\bga}{\begin{eqnarray}}
\newcommand{\eda}{\end{eqnarray}}

\newcommand{\diff}{\mathrm{d}}
\newcommand{\eps}{\epsilon}

\newcommand{\im}{\mathrm{i}}
\newcommand{\E}{\mathbf{E}}
\newcommand{\bigO}{\mathcal{O}}

\begin{document}
\title{Response of the Shockley surface state to an external electrical field:\\
A density-functional theory study of Cu(111)}
\author{K. Berland$^{1,*}$,
T.L. Einstein$^{2,\dagger}$,
and P.  Hyldgaard$^{1,*}$
}
\email[]{berland@chalmers.se}
\email[$^\dagger$]{einstein@umd.edu}
\email[$^{**}$]{hyldgaar@chalmers.se}
\affiliation{
$^1$Department of Microtechnology and Nanoscience, MC2,
Chalmers University of Technology,
SE-41296 G\"{o}teborg, Sweden\\
$^2$Department of Physics, University of Maryland,
College Park, Maryland 20742-4111, USA
}

\date{\today}

\begin{abstract}

The response of the Cu(111) Shockley surface state to an external electrical
field is characterized by combining a density-functional theory calculation for a
slab geometry with an analysis of the Kohn-Sham wavefunctions.  Our analysis is
facilitated by a decoupling of the Kohn-Sham states  via a rotation in Hilbert
space.
We find that the surface state displays isotropic dispersion, quadratic until
the Fermi wave vector but with a significant quartic contribution beyond.
We calculate the shift in energetic position and effective mass of the surface
state for an electrical field perpendicular to the Cu(111) surface; the response
is linear over a broad range of field strengths.
We find that charge transfer occurs beyond the outermost copper atoms and that
accumulation of electrons is responsible for a quarter of the screening of the
electrical field.  This allows us to provide well-converged determinations of
the field-induced changes in the surface state for a moderate number of layers
in the slab geometry.

\end{abstract}
\maketitle

\section{Introduction}

\begin{figure}[h]
\centering
\includegraphics[width=8.6cm]{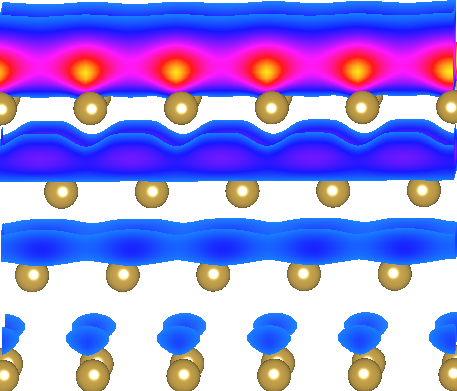}
\caption{(Color online) Contours of the $\bar{\Gamma}$-point surface-state density. Yellow
(red, blue) surfaces correspond to high (medium, low) density. The density is
largest outside the surface, and lowest in the planes formed by the atoms.
Graphics generated with VESTA.\cite{VESTA}}
\label{fig:SurfVesta}
\end{figure}

Surface states\cite{Shockley,bk:surface_states,bk:kevan,Han:prep} are long
known to significantly affect the properties of surfaces in a host of ways.
This is particularly so when the surface state crosses the Fermi level,
rendering it metallic.  Such states provide the possibility of low-energy
adsorption and enhancement of transport.  On the close-packed (111) surfaces of
noble metals, such states have their minima at the center of the surface
Brillouin zone ($\bar{\Gamma}$) in the bulk $L$ gap, have minimal angular anisotropy, and are well
approximated by free-electron dispersion.\cite{Gartland:PRB,Reinert:PRB}  Since the energy difference between
the Fermi level and the bottom of the band is small, so is the Fermi wave
vector, leading to a Fermi wavelength that is nearly an order of magnitude
larger than that of typical bulk Fermi wavelengths.  Furthermore, confinement to
the surface leads only to slow decay in directions perpendicular to the surface plane,
Fig.~\ref{fig:SurfVesta}.  Arguably
the most dramatic outcome is the observation of quantum corrals and mirages on
surfaces.
\cite{Corrals1,Corrals2,Corrals3,Corrals4}
These states can also produce ordered superstructures of adsorbed
species.\cite{surf1,surf2,surf3,bk:Einstein,surf5,surf6,surf7,surf:HighT}

A question of both fundamental and practical concern is the sensitivity of these
metallic surface states to perturbations.  Moderately strong perturbations such
as chemisorption can easily destroy the surface state,\cite{Chen:OCu,Dudde:KCu}
or create overlayer
resonances.\cite{Schiller:KCu111,NaCu111,Breitholtz:Cs_Cu111} There is also a theoretical
prediction that alkali-overlayer formation can lead to a localization (for dynamics
parallel to the surface) of high-energy electrons in a resonance state.\cite{Chis07}
On the other hand, weak perturbations can allow the state to survive but with an
altered dispersion relation (typically a shifted minimum and a changed
curvature).\cite{Carlsson:NaCu111,Schiller:KCu111}
This offers the exciting possibility to manipulate the Fermi wavelength and
effective mass of the state, if one can understand in detail how perturbations
such as adsorption affect dispersion.

However, adsorption typically leads to several different changes to the
surface.\cite{lindell:dipole}
First, there can be charge transfer, producing an electric field due to the
resulting surface dipole.  (There could also be effects from the intrinsic
dipole of an organic adsorbate.)
Second, there can be correlated electron hopping such as characteristic of
covalent bonds.
Third, the adsorbate will perturb the tails into the vacuum of the metal-surface
electron density.
\cite{lindell:dipole}
Furthermore, as dipole-producing adsorbates approach each other, they create a
depolarizing field that decreases the dipole, consistent with experiments for Na
atoms on Cu(001).\cite{Fratesi:Na_on_Cu}
Hence, the prospect of predicting how a particular adsorbate modifies the
surface state poses a considerable challenge.

In this paper, we instead focus on just the first aspect of the adsorption bond,
that due to charge transfer.
Specifically, we look at the effect that a uniform electric field normal to the
surface has on the surface state.
This problem also has advantages over a direct study of a pure ionic bond: the
electric field is uniform, and the screening is well confined to the direction
perpendicular to the substrate.
Furthermore, this simple scenario bears on the more complicated geometry
involved with the effect of strong fields between STM tips and
substrates.\cite{Stark:2003,Stark:2004}

For over four decades,\cite{LK70} with increasing sophistication, theoreticians have used density-functional theory (DFT) in various implementations\cite{LK70,WL87,CZ88,Kiejna,SurfvdW,Feibelman:Pt001,Ingles91,Negulyaev:Eswitch,Achilli} to examine the effect of static perpendicular electric fields on the electronic properties of simple metal surfaces (until relatively recently, typically jellium-like).
 Trying to
understand experiments related to self-diffusion on Pt(001), Feibelman discussed
how such electric fields can be used to tune diffusion barriers and possibly
alter the dominant mechanism of mass transport.\cite{Feibelman:Pt001}  Negulyaev
et al. showed that such fields could serve as a switching tool for magnetic
states in atomic-scale nanostructures.\cite{Negulyaev:Eswitch}  However, we know
of no DFT study of the effect perpendicular electric fields on metallic surface
states.


We further note that understanding the surface response to a
perpendicular electric field and the formation of an image plane are
fundamental building blocks in the study of the van der Waals
interactions.\cite{Lifshitz,ZarembaKohn76,Apell:Pscr,PerssonApell,PerssonZaremba,Liebsch,Tractable} The image plane is available from
DFT calculations\cite{SurfvdW,ExtSurfvdW,UnifiedvdW,Hult:free}
and the handling of screening is a central element in the vdW-DF
method.\cite{vdWDF,vdWDFsc,vdWDF2} Knowledge of the image
plane position permits an approximation of multipole
effects,\cite{ZarembaKohn76,LayervdWDF} and hence a quantitative
study of interactions of, for example, atoms and molecules at
noble-metal surfaces.\cite{ZarembaKohn77,HN:JPC,PhysPot,Kyuho:subm} The
image plane is also important for ensuring transferability over a range
of different binding distances, for example, in molecular
crystals.\cite{Kleis:nanotube,molcrys1,molcrys2}

Much of our exploration was motivated by a desire to manipulate the surface
states that mediate the interactions between anthraquinone (AQ) molecules on
Cu(111):\cite{AQ:Cu111} the giant regular honeycomb formed spontaneously
by them are likely related to such interactions.\cite{Kim:MonteCarloAQ}
Alternatively,
the pores in the honeycomb can be viewed as an array of two-dimensional (2D)
quantum dots.  The stabization of the pattern may be due to the
population---from the metallic surface state---of the 2D orbitals of the dots,
forming what amounts to closed-shell, 2D-noble-gas-like
quasiatoms.\cite{Wyrick:Do2d}
By manipulating the surface state, we hope to tune its Fermi wave vector (and,
concomitantly, its effective mass) and thereby enhance or destabilize different
superlattice structures.
A major goal of this study of surface-state response is to gain the ability to
engineer novel structures on surfaces.  Furthermore,
the standing waves within the honeycomb cells,  arising from these surface
states, are believed to determine the potentials that small molecules like CO
encounter when adsorbing within the cells. \cite{PowerOfConfinement}  These
cells form a set of identical nanostructures with thermodynamic-like behavior
that differs significantly from that found when these molecules adsorb on large
defect-free flat surfaces.

Most traditional DFT implementations rely on supercells to model surfaces,
with slabs separated by vacuum.
However, since pairs of Shockley surface states on opposite sides of
a slab hybridize, the energy and dispersion are affected, with ensuing 
loss of accuracy.
Use of very thick slabs can marginalize the hybridization effects, but
such a brute-force approach is computationally expensive and more 
susceptible
to numerical noise.  In this paper, we present a method which extracts 
proper,
un-hybridized, surface states from standard supercell-DFT studies for 
geometries
with moderate slab thicknesses.  Our method is based on a simple 
rotation in the Hilbert space spanned by the two Kohn-Sham (KS) metallic 
surface states that are found in
underlying semilocal DFT calculations.  We use our method to 
characterize the response
of the Cu(111) surface state, but it should also be useful for the study 
of other,
more complex, material systems.

The KS-rotation method presented here is complementary to the use of more
advanced DFT implementations, such as the embedding Green function
method,\cite{Inglesfield,Nekovee,Szunyogh1,Szunyogh2,Ishida,Lazarovits,Butti:ImagPot}
which effectively model a semi-infinite surface.
The embedding method can also handle an external field.\cite{Achilli}
As our study is based on a traditional DFT implementation, we do not 
correctly
describe the image-potential behavior, for which GW 
calculations\cite{Equiluz,Sau}
are normally required.  The embedded method allows an explicit
inclusion\cite{Nekovee,Szunyogh1,Szunyogh2,Butti:ImagPot} of an image-like behavior 
and can
therefore determine image-potential states in DFT.  The inclusion of this
image-like behavior would likely also improve the accuracy of the 
evanescent
part of the surface states found in the slab analysis.
 In spite of these benefits of an embedding method, we believe it is also important to continue to seek simple mechanisms to enhance the accuracy in widely used (for examples, Refs.~\onlinecite{Kokalj,Carlsson:QW,Scheffler,Bogicevic,Foelsch,Chis,Luo,Silva,Yu,Breitholtz:Cs_Cu111,Forster,Wong,Howe,Hayat,Bjork,Dyer}) slab-geometry DFT studies of surface states. 

The plan of this paper is as follows. In Sect. II we discuss our
computational methods, paying particular attention to a simple yet
unambiguous and robust way to accurately decouple the surface states
on the two sides of the slab and to obtaining a faster convergence
as a function of slab thickness.  In Sect.  III we characterize the
surface state with no electric field, while in Sect. IV we describe
the changes in the wavefunctions, potential profile, and
dispersion in the presence of an applied perpendicular field.
Sect. V discusses the screening of this electric field, deriving the relative
contribution of the electrons in the surface state.  It also makes comparisons
with experimental data and illustrates the decoupling method for benzene on Cu(111).  Finally, Sect. VI offers conclusions about the impact
of these changes in the surface state.

\section{Computational methods}

The electronic structure is obtained with DFT within the generalized-gradient
approximation (GGA) for exchange-correlation using the PBE \cite{PBE} version.
For these calculations, we use the ultrasoft pseudopotential plane-wave code
\verb DACAPO , \cite{DACAPO} with
an energy cutoff of 400 eV,\cite{Comment_density} and a \textbf{k}-sampling of
16$\times$16$\times$1.
The KS states required to obtain the surface-state dispersion is calculated
non-self consistently in a post-processing Harris functional calculation
\cite{Harris}
in which the wave vectors were sampled with a grid spacing of 0.01 $K$, where
$K$ is the size of the shortest in-plane reciprocal-lattice vector.

The Cu(111) surface is modeled as a finite slab in a supercell since the
plane-wave scheme restricts us to using periodic boundary conditions.
The top and bottom copper layers in two adjacent supercells are separated by
$12 {\rm \AA}$, thus insuring negligible cross-coupling.
The lattice constant of the slab is set to that of copper, $a= 3.65 {\rm \AA}$,
as obtained in a separate bulk PBE calculation.
The electrical fields used in this study only slightly perturb the electronic
structure of the surface, and we find no significant relaxations of the atoms of
the surface slab; the atoms are therefore frozen in their truncated bulk
positions.\cite{Comment_rel}
A dipole layer in the vacuum region induces an external electrical field normal
to the copper slab.\cite{Dipole,Neugebauer:Al111}
Both positive and negative electrical fields are studied in a single
calculation.

\label{Sec:decouple}
\subsection{Issues with coupled surface states}

The finite slab geometry makes the surface state couple to both the two sides of
the slab, and therefore challenges the analysis of surface state response to an
external perturbations.

\begin{figure}[h]
\centering
\includegraphics[width=8.6cm]{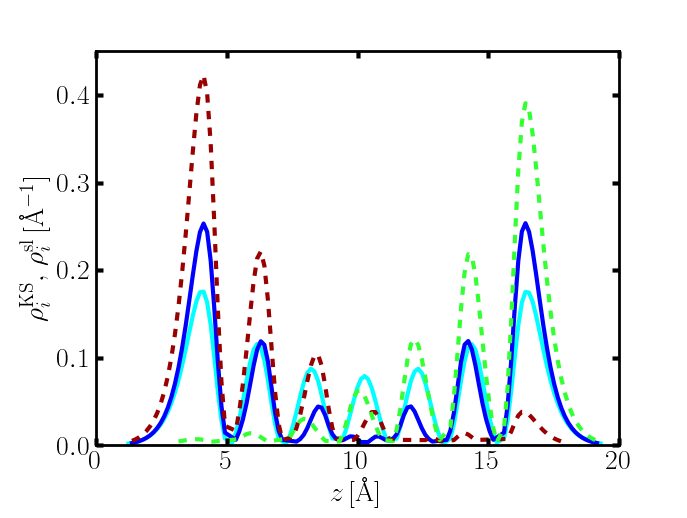}
\includegraphics[width=8.6cm]{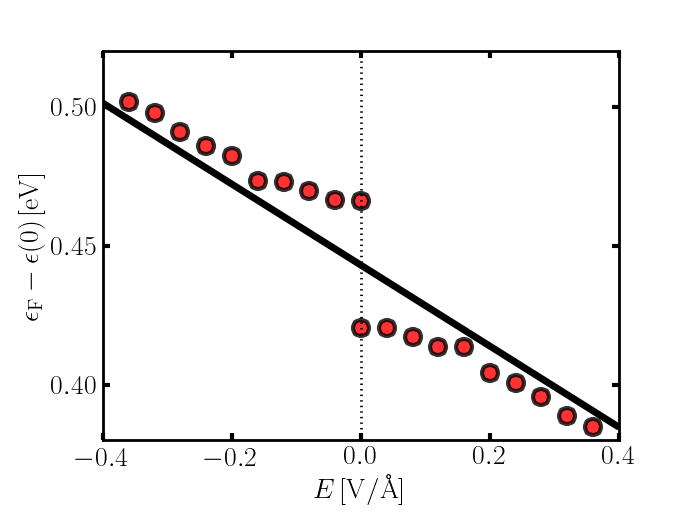}
\caption{(Color online) Issues with surface-state hybridization that arise because of the
finite slab geometry.
The upper panel shows the sum-projected surface-state densities obtained for a
six layer wide copper slab. The full lines shows the Kohn-Sham (KS) states
with surface-state character. The dashed lines show the
surface-localized (SL) states for which we here present a simple
but robust determination (since the SL states provide a more accurate
description of genuine surface-state behavior). The lower panel
shows the electric
field variation of the surface state for a 15-layer slab calculation
using the plane-wave code DACAPO.  The full line shows the
least-squares fit to decoupled surface-state energies as
obtained for the SL states, Sec. IV.B.}
\label{fig:issues}
\end{figure}

In the upper panel of Fig.~\ref{fig:issues}, the full curves show the
sum-projected density of the KS wavefunctions with surface-state character,
$\psi^{\rm KS}_i(\bk)$ at
the $\bar{\Gamma}$ point,
\begin{equation}
\rho^{\rm KS}_i(z)=\int \diff x \, \int \diff y \, |\psi^{\rm KS}_i(x,y,z)|^2,
\end{equation}
for a six-layer thick copper slab in zero external field.
These states couple equally to both sides of the slab, rather than being
localized on one side. They therefore lack the characteristic
exponential decay into the bulk.
In a hybridization, or tight-binding picture, these KS states
can be viewed as linear combinations of surface-localized (SL)
states that hybridize in a finite slab geometry.  For zero electrical
field, the KS states form symmetric and anti-symmetric combinations of
the underlying SL states (which provide an accurate description of the
actual surface-state behavior).

The dashed lines in the top panel of Fig.~\ref{fig:issues} shows the
sum-projected densities, $\rho_i^{\rm SL}$, of these SL states.
The SL states exhibit ``good" surface-state properties, like
exponential decay into the bulk and localization
at the surface. For sufficiently large slabs, these SL states
are good representations of the proper surface states, i.e., surface
states as they would have been calculated in an accurate DFT of a
semi-infinite bulk system.


In the lower panel of Fig.~\ref{fig:issues} the filled (red) circles
show, for different external fields, the KS eigenvalues for the
states with surface-state character,
$\eps_i^{\rm KS}$. These results were obtained for a 15 layer thick
slab. The jump at $E=0$ in the
KS calculation of the minimum surface-state energy, $\eps^{\rm KS}(0)$, indicates an avoided crossing, which further
supports the picture of a coupled two-level system.  Since the field influences both the properties
of the underlying states and the linear combination making up the KS states, it is a challenge to
deduce the inherent response of the proper surface states.

\subsection{Decoupling of surface states}
This subsection presents our numerically robust method to construct
the (pair of) SL states from the KS surface states of the slab.
The full line in the bottom panel of Fig.~2 shows the effective
Fermi level shifts for the SL result. The continuity of this result
contrasts the singular response which appears in an analysis based
directly on the KS states (shown as [red] dots).

Our assumption is that the KS surface states arise exclusively from
hybridization of the actual SL states; we assume that these do not couple to bulk
states or surface resonances.  In this case, we can for a given $\bk$ set up a
bonding/antibonding Hamiltonian for this SL two-level system:
\begin{equation}
\cal{H}(\bk)=\left( \begin{array}{cc} \eps(\bk) +W(\bk) & \Omega(\bk) \\
\Omega^*(\bk) & \eps(\bk)-W(\bk) \end{array} \right)
\label{eq:ls1}
\end{equation}
Here $\eps(\bk) \pm W(\bk)$ gives the energy of the uncoupled
states, where $W$ is the detuning between the levels and
$\Omega$ is the coupling. In terms of them, the eigenvalues of the KS
wavefunctions are given by
\begin{equation}
\eps^{\rm KS}_{1,2}=\eps \pm \sqrt{W^2+|\Omega|^2}\,,
\label{eq:ksFromsl}
\end{equation}
which follows from the invariance of $\mathrm{Tr}\, \cal{H}$ and $\mathrm{Det}\,
\cal{H}$.
As the slab size increases, $\Omega\rightarrow 0$ and the eigenvalues of the
decoupled surface states reduce to $\eps \pm W$.

The key observation is that the SL state energy $\eps$
converge much faster as the slab size increases than the
coupling parameter $\Omega$ vanishes.
The same is generally true of the detuning parameter $W$; even
at $E=0$ (when the correct detuning must vanish identically), we
find (below) that the numerical value of $W$ converges fairly well
before the accumulation of numerical noise
eventually destroys this behavior.  Therefore, by constructing
$\cal{H}$ of Eq.~(\ref{eq:ls1}), we can characterize
the surface-state properties to high accuracy, with a smaller unit cell
than what is required for the KS to decouple due to asymmetry induced by
small-scale numerical effects.

In general a SU(2) rotation is needed to construct the SL basis states
and $\cal{H}$ as we seek a linear combination of the KS wave-functions
$\psi_i(\bk)$ that localizes the state on one side of the slab through
constructive and destructive interference.  However, when we can
identify the symmetry point of the unit cell in the xy-plane, we can ensure that both
KS states have the same variation in the complex plane.
This is done, for example, in the present Cu(111) slab study, by
setting the KS states real at this symmetry point via multiplication
by a simple phase factor: $\psi_i \rightarrow \exp(\im \eta) \psi_i$.
The transformed KS states must then have the same complex-phase variation
across the unit cell because the KS states are Bloch functions: they
can be written as $\psi_i(\br) = u_i(\br) e^{\im \bk \br} $,
where $u_i(\br)$ has the same periodicity as the supercell and
can be chosen real.\cite{Kohn:Analytical}

With this transformation, the full SU(2) transformation reduces
to an O(2) rotation $R$ in the Hilbert space:
\begin{equation}
\left(\begin{array}{c} \phi_1 \\ \phi_2\end{array} \right) = \left(
\begin{array}{cc} \cos\theta & -\sin\theta \\ \sin\theta & \cos\theta
\end{array}\right)\left(\begin{array}{c} \psi_1 \\ \psi_2\end{array} \right)\,.
\end{equation}
To determine the value of $\theta$, we choose a condition of maximal
localization (MaxLoc), that is, we minimize the sum of the variances of the generated wavefunctions
\begin{equation}
F(\theta)= \sum_i |\langle \phi_i | z_i^2 | \phi_i \rangle| - |\langle \phi_i |
z_i | \phi_i \rangle|^2\,.
\label{eq:MaxLoc}
\end{equation}
Such a condition has also been  used to construct the generalized-Wannier
functions. \cite{MLW}

The matrix transformation that corresponds to this rotation is $H \rightarrow
RHR^{\rm T} = H_{\rm SL}$. Thus,
\begin{equation}
 \left( \begin{array}{cc} \eps_1^{\rm KS}  &0  \\ 0 & \eps_2^{\rm KS}
\end{array}\right)\rightarrow
\left( \begin{array}{cc} \bar{\eps}^{\rm KS}+\frac{1}{2}\Delta \eps^{\rm KS}
\cos 2\theta&
 \frac{1}{2}\Delta \eps^{\rm KS} \sin 2\theta \\ \frac{1}{2}\Delta\eps^{\rm KS}
\sin 2\theta & \bar{\eps}^{\rm KS}-\frac{1}{2} \Delta \eps^{\rm KS}\cos 2\theta
\end{array}\right)\,.
\end{equation}
Here $\bar{\eps}^{\rm KS}=(\eps_1^{\rm KS}+\eps_2^{\rm KS})/2$ and
$\Delta \eps^{\rm KS}=(\eps_1^{\rm KS}-\eps_2^{\rm KS})/2$
By comparing to Eq.~(\ref{eq:ksFromsl}), we identify the SL-hybridization parameters,
$\eps=\bar{\eps}^{\rm KS}$, $W=(\Delta \eps_{\rm KS}/2) \cos(2\theta)$, and
$\Omega=(\Delta \eps_{\rm KS}/2) \sin(2 \theta)$. We note that $\Omega$ is
real because we can here work with an O(2) rotation.

For perfectly symmetric surfaces we have $\theta=\pi/4$, and the surface-state
energy is then the average of the eigenvalues of the two states. Thus, the full
machinery discussed here is not required.
For certain perturbations, like adsorbate systems, a less elegant
solution is to use symmetric adsorbates on both sides of the slab.
Such a brute-force approach has the drawbacks of increased
computational costs and fewer layers with
a bulk-like behavior; furthermore, wave
functions are not decoupled.  Strong asymmetric perturbations,
such as halogen overlayers, lead to $W \gg \Omega$ and a
natural decoupling of the states. However, a natural decoupling
does not happen for weak perturbations, like adsorbates bound by
van der Waals interactions,\cite{benzCu} or for dilute chemisorbed
overlayers, where $\Omega \sim W$ unless a huge number of layers
are used, typically beyond the computational feasibility.

\subsection{Convergence of surface-state properties}
\label{sec:conv}
Our method to construct the SL states significantly reduces the
number of layers needed to accurately characterize the surface state;
however, the slab still must be thick enough to describe most of the relatively
slow decay into the bulk.  We also confirm that the MaxLoc condition (see
Eq.~\ref{eq:MaxLoc}) properly decouples the two states by comparing with the results
for a 24 layer calculation, where decoupling arises numerically as
$\Omega\rightarrow 0$.

\begin{table}[h]
\begin{ruledtabular}
\caption{Convergence of the surface state with slab thickness. All three columns
of energies are in meV.  The bold numbers, for 15 layers, are used for the rest
of this paper.}
\begin{tabular}{cccc}
\label{tab:conv}
\# layers & $\eps_{F}-\eps(0)$    &  $\Omega^{(E=0)}  $   & $W^{(E=0)} $\\
6   & 520   & 272  &  $0.24 $   \\
9   & 464   & 113  &  $0.38 $  \\
12  & 449   & 50  &   $0.17 $ \\
{\bf15}& {\bf 443} & {\bf 23}  &  ${\bf 0.81}$ \\
 18 & 443   &  10 & $1.34 $  \\
24  &  442  &  2 & $11.0$ \\
\end{tabular}
\end{ruledtabular}
\end{table}

Table \ref{tab:conv} shows the calculated SL state parameters
for for different slab thicknesses at zero electrical field.
The Fermi energy $\eps_{\rm F}$ relative to $\eps(0)$ converges to the sub-meV
level for a 15-layer slab if the decoupling method is used.
For a six-layer slab the surface-state energy differs from the converged value
by 80 meV, or about 20\%; we consider it a minimum slab thickness for an
approximate account of the surface state, which can be useful for studying
surface-state shifts for adsorbates-systems requiring a large supercell in the
in-plane direction.

We find that coupling $\Omega$ decays significantly slower than the value
of $\eps_{F}-\eps(0)$ converges, a fact which as mentioned above
motivates our approach. We note that the value of 272 meV for six
layers will almost deplete one of the two surface-related KS bands
(while driving the other at least partially into the energy range of
bulk states).  Nevertheless, the corresponding result for the effective
depth of the surface-state, Fermi see, $\eps_{F}-\eps(0)$ differs from the
converged value by merely 57 meV.
For 15 layers $\Omega$ reduces to 23 meV, while the $\eps_{F}-\eps(0)$
is converged to within less than an meV.

\subsection{Effects of numerical noise on optimal slab geometry}

The nonzero value of the detuning $W$ at zero electrical field stems
purely from numerical noise and grid effects, since the slab geometry
is symmetric.
The upper panel of Fig.~\ref{fig:wf} shows the KS states for different
electrical fields.  For zero electrical field, the slight asymmetry of the
curves hints of the nonzero $W$.
For larger fields, the KS states clearly favor one side of the slab, but even
for $E=0.36 {\rm V/\AA}$, the wavefunctions are localized on both sides of the
slab.
The inserts illustrate (purple dot) the coordinate in the 2D space $(W,\Omega)$,
with magnitude $(\eps^{\rm KS}_1-\eps^{\rm KS}_2)/2$, constructed using the MaxLoc
condition.
The detuning grows as the electrical field increases, while the coupling remains
roughly constant.

For six layers the two KS surface-state densities are fully symmetric
as shown in the upper panel of Fig.~\ref{fig:issues}. For even wider
slabs, the coupling
reduces, and the surface-state densities becomes increasingly
asymmetric; for 24 layers, we find that the surface states
localize almost exclusively on one side of the slab, consistent
with a detuning five times larger than the coupling
$\Omega$.

That $\eps^{\rm KS}\approx \eps$ for 24 layers reflects this
finding and the general expectation that increasing the number of
layers must eventually decouple the KS and ensure an automatic
approach to the more meaningful SL states.  Nevertheless, we find
that the increasing value of $W$ for slabs with than 15 layers
is the sole reason for this decoupling. As this increase in
$W$ (for $E=0$) is solely an expression of an accumulation
of numerical noise (with the number of atoms and electrons),
we conclude that such brute-force decoupling is not desirable. 

We propose instead to use the maximum-localization approach to
identify the genuine surface-state behavior and avoid such
numerical noise. In this study we use 15 layers, as
it is an optimum between a converged value of $\eps$ and a
minimal numerical noise; our approach converges the surface-state
energies to meV accuracy.

\section{Surface state dispersion in zero electrical field}

Our calculations of the surface-state dispersion in zero electrical field are
presented in this section and compared to experimental\cite{exp:STM,Kevan}  and
earlier theoretical studies. \cite{surf5,Cu111:relLDA}
We find that the surface state is practically isotropic even for $k\equiv|\mathbf{k}|>2.5k_F$, and
that non-parabolicity becomes significant once $k$ exceeds $k_F$.

\begin{figure}[h]
\centering
\includegraphics[width=8.6cm]{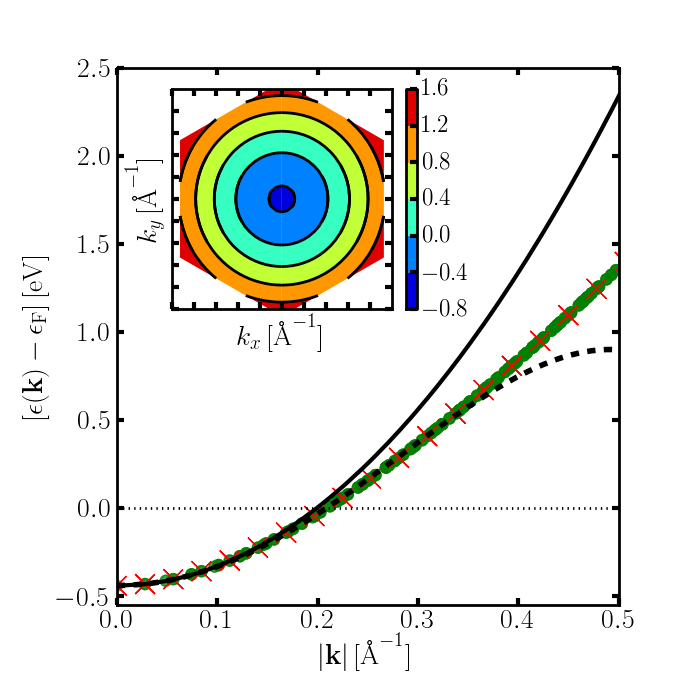}
\caption{(Color online) Dispersion of the surface state plotted as a function of $|\mathbf{k}|$
for ${\mathbf{k}}$ sampled evenly on a 2D grid.
The full (dashed) curve gives the best-fit second- (fourth-) order polynomial.
The dotted line indicates the Fermi surface.
The filled [green] circles, which form a thick line, indicate the calculated
values, while the [red] crosses indicate those for $k_y=0$. The insert shows the
energy contours as function of $\mathbf{k}$, with ticks having the same spacing
as in the main figure.}
\label{fig:Disp}
\end{figure}

Figure~\ref{fig:Disp} shows the surface-state dispersion as a function of the
absolute value of $\mathbf{k}$.
The filled (green) circles indicate the calculated values of
$\eps(\bk)-\eps_{F}$, which have been sampled evenly on a $k_x,k_y$-grid, while
the (red) crosses indicate the energies obtained for $k_y=0$.
That they align to form a curve shows that the surface state is isotropic, even
as far as $\sim 2.5 |\mathbf{k}_F|$.
This isotropy is further evidenced by the circular constant-energy contours
displayed in the insert.
The full curve gives the parabolic dispersion, while the dashed includes
non-parabolicity via a quartic term:
\begin{equation}
\epsilon(k)-\eps(0)=\frac{\hbar^2 k^2 }{2 m} - \alpha k^4 +
\bigO(\bk^6)\,,
\end{equation}
with parameters obtained as described in the following paragraph. These curves
show that up to about the $k_{F}$, the dispersion is well described by a
parabolic form, but for larger wave vectors, non-parabolicity is significant.

The deviation from a parabolic-dispersion behavior has been observed experimentally and
can be understood in terms of an $s$-band tight-binding model.\cite{Kern:devParabolic}
We expand $\epsilon({\bk})$ as given by the standard Hamiltonian and
find:
\begin{align}
\hspace{-10mm}\lefteqn{\epsilon({\bk})-\eps(0) \propto 1 -\frac{1}{3} \cos(k_x
a) -\frac{2}{3}\cos\left(\frac{k_x a }{2}\right)\cos\left(\frac{\sqrt{3}
k_ya}{2}\right)} \hspace{5mm} \nonumber  \\
&&=  \frac{1}{4} (ka)^2 - \frac{1}{64} (ka)^4 + \frac{10 + \cos(6
\theta)}{23040}(ka)^6 + \bigO(k^8) \, .
\label{eq:sl}
\end{align}
\noindent where $\theta \equiv \arctan(k_y/k_x)$ and $a=2.58{\rm \AA}$ is the
nearest-neighbor distance.   Thus, the quartic correction is negative (albeit much smaller than found
in the DFT calculation), and
anisotropy does not appear until the $k^6$ term.\cite{subPatrone}

Table \ref{tab:prop} gives deduced surface-state properties and compares them to
other studies.
In obtaining these values, we only needed data points obtained for $k_y=0$,
since we have amply demonstrated the isotropic surface-state dispersion; this
procedure also
avoids excessive weighting of large-$|\mathbf{k}|$ values, as the number of 2D
grid points grows approximately linearly with the wave-vector magnitude.
Our values for the effective Fermi level is similar to earlier experimental and calculated values. Our value is somewhat smaller than that given in Ref~\onlinecite{Butti:ImagPot}, which is based on a semi-infinite approach including an image potential. Our value for the effective mass is smaller than those of the earlier studies, which is partly a result of the fitting procedure used to obtain its value.

A parabolic fit is often used to extract the effective mass from the dispersion curve; however, the effective mass is defined by the second-order Taylor expansion of the dispersion curve, not by an optimal parabolic fit to a curve that may deviate from parabolicity.
To properly extract the terms in the Taylor expansion with a polynomial fit, we
rely only on values fairly close to the $\bar{\Gamma}$~point and, in addition, include higher-order terms in the fit to minimize their influence on the extraction of the lower-order ones. It is possible to avoid the influence of quartic terms on the effective mass with a purely parabolic fit, but this requires that we restrict the domain to data points very close to the $\bar{\Gamma}$~point. This has the disadvantage that the influence of noise is larger.
In detail, our approach is as follows: first, the effective mass $m$ is obtained with
least-squares using a fourth-order polynomial fit for the six smallest
$\mathbf{k}$ points (corresponding to 6\% of the
reciprocal vector); next, we keep the mass fixed and
determine $\alpha$ using a
sixth-order polynomial using all 20 data points.

The correspondence with the data points of
Fig.~\ref{fig:Disp} for small and medium $|{\bk}|$ corroborates
our procedure, as just described.  If we instead include ten data points and fit
to a purely parabolic dispersion, we find a mass of 0.38$m_e$,
closer to that of earlier studies (listed in Table \ref{tab:prop}).
Thus, that our deduced mass is smaller than
previously obtained values relates to our use only of
data points close to the $\bar{\Gamma}$ point and our
inclusion of higher-order polynomial terms.
The quartic prefactor $\alpha$ compensates somewhat
for the smaller mass, and the Fermi wavelength $\lambda_F$
and wave vector $k_F$ are in good agreement with earlier results.
That sensitivity to fitting domain (parabolic fitting) was discussed in Ref~\onlinecite{Butti:ImagPot}.
They note that when comparing with experimental data, it is important that same procedure is used in both cases.
We argue that, ideally, all parameters should be defined in terms of the Taylor expansion.
When they use a minimal sampling of $\bk$ points around the $\bar{\Gamma}$~point, they obtain an effective mass of $0.303 m_e$, which is closer to, and even smaller than the mass we obtain.
The importance of making sure that higher-order terms do influence the
extraction of the effective mass is also reflected in the significant non-parabolic dispersion that Becker et al.\cite{Berndt:Ag111} found for Ag(111).
They extracted the effective mass by considering only data point collected close to the $\bar{\Gamma}$~point.

\begin{table}[h]
\begin{ruledtabular}
	\caption{Properties of the surface state.
	The value $\eps_{F}-\eps(0)$ is the difference
	between the Fermi surface and the minimum of the surface
	state $\eps(0)$. Also listed are comparisons of
	our values of the effective mass, surface-state Fermi
	wavelength and wavevectors, $m, \lambda_F, k_{F}$.  Our
	DFT value for the quartic component of the in-plane surface
	state dispersion, $\alpha$, is compared with the
	($m$-dependent) estimates $(\hbar a)^2/32 m\approx 1.58 (m_e/m){\rm eV{\AA}^4}$ given by
	the expansion (\ref{eq:sl}) of
	the tight-binding behavior observed in
	Ref.~\onlinecite{Kern:devParabolic}.}

\label{tab:prop}
\begin{tabular}{lllll}
                & Here & Other theory  & STM  &  ARPES   \\
		$\eps_{F}-\epsilon(0)[{\rm eV}] $  & 0.443  &  0.42\footnotemark[1], 0.40\footnotemark[2], 0.526\footnotemark[6]  &
0.42\footnotemark[3]  & 0.39\footnotemark[4]        \\
$m/m_e $                  &   0.34 &   0.38\footnotemark[1],
0.43\footnotemark[2], 0.394\footnotemark[6]     & 0.38\footnotemark[3]    &  0.44\footnotemark[4]    \\
$\lambda_F [{\rm \AA}]$   &   30.2          &    31.0\footnotemark[1]       &
30.0\footnotemark[3]       &        \\
$k_F [{\rm \AA}$          &     0.208        &    0.20\footnotemark[1]      &
0.21\footnotemark[3]     &        \\
$\alpha [{\rm eV \AA^4}] $  & 23.1  & 3.7-4.7\footnotemark[5]
&  & \\
\end{tabular}
\footnotemark[1]{Ref.~\onlinecite{surf5}}
\footnotemark[2]{Ref.~\onlinecite{Cu111:relLDA}}
\footnotemark[3]{Ref.~\onlinecite{exp:STM}}
\footnotemark[4]{Ref.~\onlinecite{Kevan}}\\
\footnotemark[5]{Tight binding, Eq.~(\ref{eq:sl}).}
\footnotemark[6]{Ref.~\onlinecite{Butti:ImagPot}}
\end{ruledtabular}
\end{table}

\section{Surface state in an external electrical field}

The dispersion of the Shockley surface state was characterized in the previous
section at zero electrical field. In this section, we present the results for a
finite external electrical field.

\subsection{Wavefunctions and potential profile}

Figure~\ref{fig:SurfVesta} gives intersections or contours of the
variation of the surface-state density in three dimensions.
Fig.~\ref{fig:SurfVesta} confirms that the surface state is
mostly located between the copper layers or outside the outermost
copper layer. The figure also shows that, in the
in-plane direction, the density is largest close to
the copper atoms.

\begin{figure}[h]
\centering
\includegraphics[width=8.6cm]{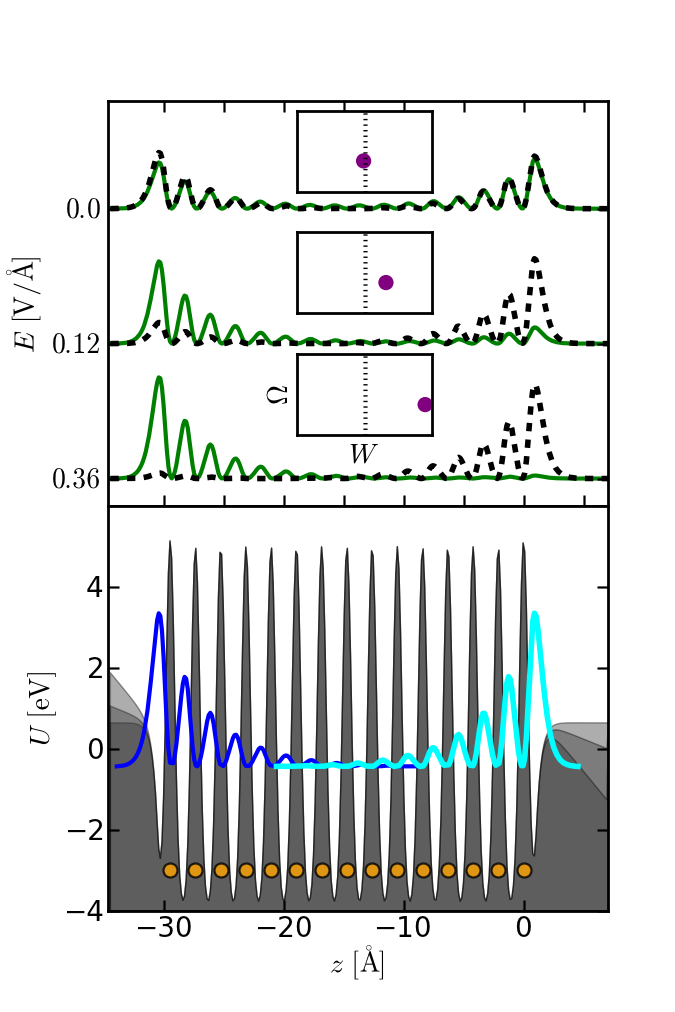}
\caption{(Color online) The average potential profile and KS and SL surface-state densities at
the $\bar{\Gamma}$ point for a Cu(111) surface in an external electrical field $\mathbf{E}$
pointing to the left, i.e., inwards to the surface at the top
(at $z\approx 0$) of the slab and outwards from the surface at the bottom (at
$z\approx -30\hbox{\AA}$).  Thus, the force on an electron is away from the
surface at the top of the slab and the minimum energy $\eps(0)$ of this
top-surface SL state is lowered, increasing $\eps_F-\eps(0)$.
The upper panel shows the KS surface-state densities (full and
dashed) for three different strengths of \textbf{E}.
At zero electrical field they are almost symmetric, while for larger field they
almost decouple. The inserts illustrate the corresponding hybridization
parameters. The lower panel shows the atom positions, filled (yellow) circles,
and the decoupled SL surface states, evanescent oscillatory curves.
The bottom panel also shows, by surfaces of semi-transparent grey shading,
the potential profile for the three different strengths of electrical fields.
Field-induced changes in the potential profile are observable only outside
the outermost Cu atoms.}
\label{fig:wf}
\end{figure}

The upper panel of Fig.~\ref{fig:wf} shows the sum-projected density of the KS
wavefunctions at the $\bar{\Gamma}$ point, $\rho_i^{{\rm KS, (}E{\rm )}}(z)$, for
different external electrical fields $E$.  For zero field (top line), the
two wavefunctions form approximately symmetric and anti-symmetric
functions as depicted by the full (green) and dashed (black)
curve, respectively.  That the two curves have almost equal magnitude on the left side,
but not on the right, reflects the presence of numerical noise and
grid sensitivities in the plane-wave DFT calculations.

The lower panel of Fig.~\ref{fig:wf} shows the sum-projected density of the
SL states at the ${\bar \Gamma}$ point, $\rho_i^{{\rm SL, (}E{\rm )}}(z)$,
  and the average potential energy as a function of $z$.
The SL states differ marginally for different electrical fields; hence, we only
display those for zero electrical field.
These states exhibit oscillatory exponential decay into the bulk and are located
solely on one side (or the other) of the slab, further validating use of the MaxLoc
condition.
The filled (yellow) circles indicate the location of the copper atoms.
The first node of the wavefunctions coincides with the position of the top
copper layer.  The (laterally averaged) potential energy profiles for the three
field strengths are given by surfaces of semi-transparent grey shading; thus,
the light-,medium-, and dark grey areas indicate energies which are smaller
than one, two, or all three of the potentials. A sharp edge between dark grey and
white indicates that the potentials overlap, for example within the slab.
At the left side of the slab, the finite electrical fields go
outward from the surface, corresponding to a total reduction of
electrons.  The number of surface-state electrons are also reduced
on the left side as the electrical field push the electrons toward
the surface and thereby increases confinement. This raises the minimum
energy $\eps(0)$ of the surface state, lowering $\eps_{F}-\epsilon(0)$.
At the right side, electrons accumulate and so do the surface state
electrons; they experience weakening confinement, hence
a lowering of the energy which leads to an increasing value of
$\eps_{F}-\epsilon(0)$.

For large electrical fields, the surface state becomes increasingly unstable,
and at some point, depending on the size of the vacuum region in the
unit cell,
electrons start accumulating at the dipole layer, ruining the physical picture
of semi-stable surface states.

\subsection{Modified dispersion}
\label{moddisp}

\begin{figure}[h]
\centering
\includegraphics[width=9cm]{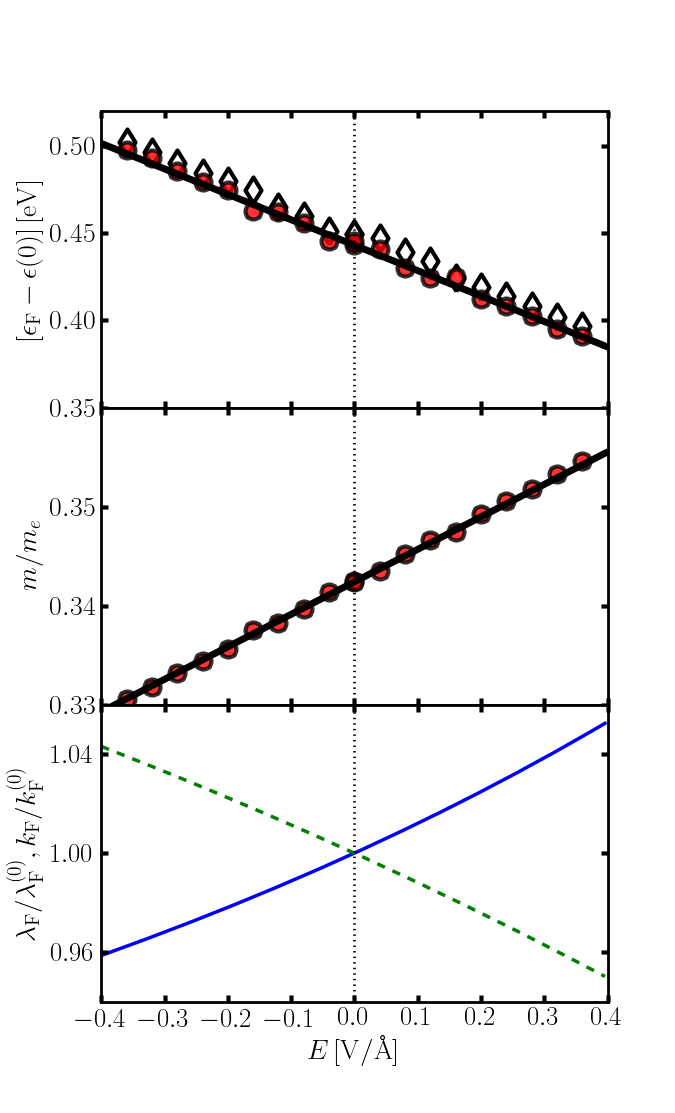}
\caption{(Color online) Calculated energy at the ${\bar{\Gamma}}$-point (upper panel) and mass
(middle panel) shift for different external electrical fields. Here we take a positive value
of the field strength $E$ to imply that the field $\mathbf{E}$ (the force) points away from
(towards) the surface. In terms of the underlying slab calculations, Fig.~\ref{fig:wf}, the
positive-$E$ (negative-$E$) results characterize the behavior of the SL state at the left-most
(right-most) slab surface. The filled (red) circles [(white) diamond] gives the
calculated values for a 15 [12] layer slab. The black line gives the least-square fit to the
calculated data. In the lower panel the relative shift in wavelength (wave
vector) is given by the full (dashed) curve.}
\label{fig:shift}
\end{figure}

The external electrical field not only modifies the effective Fermi level
$\epsilon_{F}$ of the surface state, but also influences the
surface-state dispersion, expressed here in terms of a shift
in the effective mass.  The mass is obtained as described in
the previous section.

Figure~\ref{fig:shift} shows the calculated values of the surface-state energy
relative to the effective Fermi level (upper panel), the effective mass (mid
panel),
and in the lower panel, the relative shift in the wavelength and the wave vector
based on the shift in Fermi level and effective mass in the parabolic
free-electron gas approximation.
It shows that for the largest plotted electrical field the wavelength shifts by about $4
\%$.
For the upper and mid panel, the line gives the least-square fit according to
linear relations:
\begin{eqnarray}
\left[\eps_{F}- \eps(0)\right]_{E}& =&  \left[\eps_{F}- \eps(0)\right]_{E=0} +
A_{\eps } E   \\
 m^{(E)}/m_e &=&  m^{(0)}/m_e + A_{\hat{m}} E  \, , \label{eq:relations}
\end{eqnarray}
where $\hat{m}$ denotes $m/m_e$.

The parameters determined are listed in Table~\ref{tab:Efield} along with the
shift in the characteristic Fermi wave vector $k_{F}$ ($ A_{\rm k_{F}}$) and in
the wavelength $\lambda_{F}$ ($A_{\lambda_{F}}$).
\begin{table}[h]
  \caption{Response of the surface state to the external electrical
  field, obtained from fitting to the calculated dispersion.  The values
  $A_{\eps}, A_{\hat{m}}, A_{\lambda_{F}}$, and $A_{k_F}$ are
  our results for the field derivatives of the surface-state
  Fermi sea level, $\eps_F-\eps(0)$, effective mass, Fermi wavelength, and
  Fermi wavevector, respectively. The value of $A_{\sigma}$ is our
  DFT-based determination of the fraction of overall external-field
  screening which originates from the surface-state behavior.}
\label{tab:Efield}
\begin{tabular}{@{\hspace{5mm}}l@{\hspace{40mm}}r@{.}l@{\hspace{5mm}}}
\hline \hline
$ A_{\eps} [{\rm e \AA} ]$ & -0&146 ($\pm 0.007$)\\
$ A_{\hat{m}} [{\rm \AA/V}]$  &  0&029 \\ 
$ A_{k_F}  [{\rm V}^{-1} ] $ & -0&025 \\
$ A_{\lambda_{F}} [{\rm \AA}^2 {\rm V}^{-1} ]$  & 3&68 \\
$A_\sigma  [{\rm (V \AA)^{-1}}]$  & -4&$82\cdot 10^{-3}$   \\ 
\hline \hline
\end{tabular}
\end{table}

The range of field strengths $E$ plotted in Fig.~\ref{fig:shift} are larger than
values which are directly achievable in most experiments (but smaller than what is
regularly achieved in adsorption studies, as discussed in Sec.~V). In particular, in a study of the Stark
shift of the surface state due to measurement by scanning tips rather than
photoemission, Berndt's group\cite{Stark:2004} found a linear decrease in
$\left[\eps_{F}- \eps(0)\right]_{E}$ vs.\ $E$. This is true up to
$E$ = 0.055 V/{\AA}, at which there is a energy shift of magnitude
$\sim$13 meV; thereafter, the rate of change increased
markedly, an effect which the authors attributed to the
breakdown of the tunneling regime at small tip-surface separation.
We note that while the experimentally observed breakdown occurs around
0.055 V/{\AA} for Cu(111), it occurs already at 0.008 V/{\AA}
for the more weakly bound Shockley state on Ag(111).

From their graph of energy shift vs. field, we
extract the mean slope to be
$A_{\eps} \approx -0.23$ e{\AA} for Cu(111), comparable to
our computed value in Table \ref{tab:Efield}.  We speculate that
the larger magnitude in the experiment
arises from the additional strain in lateral directions due to the
non-uniform field between the STM tip and the surface.

\section{Discussion}
\subsection{Role of the surface state in screening an external
field}

In a metal, the electrons close to the Fermi level completely screen an external
static electrical field perpendicular to the surface.
The virtually identical potential profile inside the slab, for $-30 {\rm \AA}
<z<0$, in the lower panel of Fig.~\ref{fig:wf} for different electrical fields
shows that our DFT calculations accounts for this screening.

\begin{figure}[h]
\begin{center}
\includegraphics[width=8.6cm]{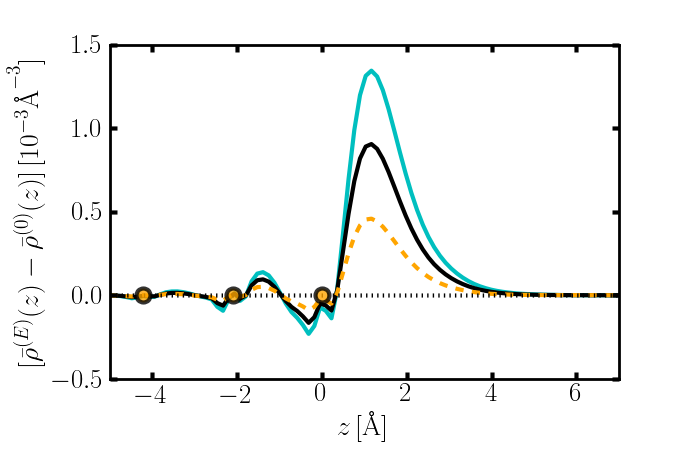}
\end{center}
\caption{(Color online) Charge transfer induced by external electrical field averaged over the
in-plane directions. The upper light (cyan) [dark (black), dashed (orange)] is
for $E= -0.36\, [-0.12, 0.24]\, {\rm V/\AA}$.}
\label{fig:screening}
\end{figure}

Figure \ref{fig:screening} displays the charge density response to the external
fields. The charge is induced outside the outermost copper atom, while there are
small oscillations into the first two bulk layers. The response is almost
identical, with opposite prefactor, at the other side of the slab.
The short screening length indicates that the bulk electrons play a major role
in screening the electric field.

To determine the fraction of the screening performed by the surface state
compared to the bulk states, we obtain the total charge induced on the surface
from Gauss's law and the charge accumulated in the surface state with its
parabolic free-electron dispersion.
Gauss's law relates the induced charge on the surface to the electrical field by
$\sigma=\varepsilon_0 E$.
To obtain the charge accumulation in the surface state, we recall that
2d-electron density of states (including spin degeneracy) is $m \eps_{F}/\pi$
(assuming in this paragraph, for notational simplicity, that  $\eps(0) = 0$).
Then the charge density due to the
surface state is $\sigma_s= -e m \eps_{F}/\pi$ at $T=0$, which is the
appropriate temperature for comparison with standard DFT calculations.
Since the mass $m$ and the Fermi-level $\eps_{F}$  change as an external
electrical field is applied, so does the charge of the surface state. To first
order in the electrical field, this shift is given by
$\Delta \sigma_s= - \Delta (m \eps_{F}) e / \pi =- A_\sigma e E/\pi+\bigO(E^2)$,
where $A_\sigma= \left(m A_{\eps} + A_{\hat{m}} \eps_{F}\right)$.
Combining the expressions, we find the fraction of the electrical field screened
by the surface state:
\begin{equation}
  \frac{\Delta \sigma_s}{\sigma} = \frac{- e A_{\sigma}}{\varepsilon_0 \pi} +
\bigO(E) = 0.27\,.
\end{equation}
The surface state therefore plays a significant but not dominant role in the
screening of the electrical field.  We note that the calculated potential profile and
induced charge density curves exhibit minute oscillatory variation for different
fields. This observation, again, accords well with the idea that bulk electrons perform
most of the screening.

That screening arises predominantly from bulk electrons has several
noteworthy consequences.
It supports the practice of calculating interaction energies due to
surface-state mediated interactions\cite{surf2,surf1,surf3,surf5,surf7} based
solely on the changes in the energies of single-particle states,
ignoring the electrostatic term in the full Harris formalism.\cite{Harris}
It
also suggests that interactions between closely spaced adsorbates (within a few
lattice spacings of each other) are dominated by the bulk states rather than the surface states.
Based on a STM study Petersen et al.\cite{Petersen} concluded that screening of step edges at the surface was dominated by the surface states on Cu(111) an Au(111), but they also noted that screening of defects slightly below the surface is dominated by bulk electrons.
The important role of the bulk electrons is also reflected in the linear dispersion of acoustic-surface plasmons on Cu(111).\cite{asp1,asp2}

\subsection{Comparison with experiments}

\begin{table}[h]
\begin{ruledtabular}
\caption{Experimental results for alkali adsorbates and electric fields on the
Shockley surface state on Cu and Ag~(111). The ratio of effective to bare
electron mass is indicated by $\hat{m}$. Unless indicated otherwise, coverages
are a saturated overlayer (``1 ML").  The entries for electric fields, denoted
by ``E:substrate", are those discussed in Subsection \ref{moddisp}; the
``adsorbed" values are for the the maximum fields before the tunneling regime
breaks down (not the tabulated\cite{Stark:2004} values for $R=500 M\Omega$).}
\begin{tabular}{lccccc}
\label{tab:alkss}
System & \multicolumn{2}{c}{[$\eps_F -\eps(0)]$ [meV]}  &
\multicolumn{2}{c}{$\hat{m}$} \\
& clean & adsorbed & clean & adsorbed   \\
$\lesssim \frac{1}{2}$ML Li/Cu \footnotemark[1]${}^,$\footnotemark[2] &   & 755  & & \\
0.11-0.15ML Na/Cu \footnotemark[3] & 390  & 800  & & \\
0.4ML K/Cu \footnotemark[4] & 410 & 755 & 0.41 & 0.36 \\
Cs/Cu \footnotemark[1] &   &   & & \\
Ar/Cu \footnotemark[5] & 434(2) & 376(3) &0.43(1) & 0.46(3) \\
Kr/Cu \footnotemark[5] & 434(2) & 358(2) &0.43(1) & 0.44(2) \\
Xe/Cu \footnotemark[5] & 434(2) & 291(2) &0.43(1) & 0.44(2)\\
Xe/Cu \footnotemark[6] & 440(10) & 310(10$^+$) &0.40(2) & 0.42(3)\\
$\frac{1}{4}$ML Na/Cu \footnotemark[7] & $\sim$340 & 690 & & \\
C$_6$H$_6$/Cu  \footnotemark[8]  & 410  & 240  & 0.46 & 0.9  \\
E:Cu \footnotemark[9] & 437(1) & 450 &  &\\ \hline
0.17ML Na/Ag \footnotemark[10] & -65\footnotemark[11] & -260 & &  \\
Ar/Ag \footnotemark[5] & 62(2) & -1(3) &0.42(1) & 0.46(4) \\
Kr/Ag \footnotemark[5] & 62(2) & -08(2) &0.42(1) & 0.44(4)  \\
Xe/Ag \footnotemark[5] & 62(2) & -52(2) &0.42(1) & 0.42(6) \\
Xe/Ag \footnotemark[12] & 67 & -52 &\multicolumn{2}{c}{ratio = 1.00(15)} \\
E:Ag \footnotemark[13] & 64(1) & 71 & &  \\
\end{tabular}
\footnotemark[1]{Ref~\onlinecite{Carlsson:LiCu111}
 indicates that the surface state shifts down with Li adsorption and disappears (into the bulk) before a half monolayer.}
\footnotemark[2]{Ref~\onlinecite{Kliewer:NaCu111}}
\footnotemark[3]{Ref.~\onlinecite{Carlsson:NaCu111}}
\footnotemark[4]{Ref.~\onlinecite{Schiller:KCu111}}
\footnotemark[5]{Ref.~\onlinecite{Forster:SurfInt}}
\footnotemark[6]{Ref.~\onlinecite{Park00}}
\footnotemark[7]{Ref.~\onlinecite{Wallden80}}
\footnotemark[8]{Ref.~\onlinecite{munakata:2736}}
\footnotemark[9]{Ref.~\onlinecite{Stark:2004}}
\footnotemark[10]{Ref~\onlinecite{Carlsson96Lett}}
\footnotemark[11]{Ref~\onlinecite{Wang2011}}
\footnotemark[12]{Ref.~\onlinecite{HGR01}}
\footnotemark[13]{Ref.~\onlinecite{Stark:2003}}
\end{ruledtabular}
\end{table}

In Table \ref{tab:alkss} we collect results for adsorption-induced shifts in
$\eps_F -\eps(0)$, as well as changes in the effective mass and the workfunction
when available, for Cu(111) and Ag(111) to provide information about the kinds of values measured in
mostly-recent experiments.  We have not included gold-surface-adsorbate systems because of the strong spin-orbit coupling gives rise to a Bychkov-Rashba  splitting.\cite{Bychkov-Rashba,LaShell,Simon2011} In addition, Au(111), in contrast to Cu, Ag, and other fcc (111) surfaces, reconstructs, taking on a herringbone pattern.\cite{Lippel1989,Barth1990}
We note that is not always possible to fully calibrate the
values in the experiments since precise coverages are rarely given.  Data is typically
for a monolayer (1 ML), which refers to the saturation coverage rather than one
adsorbate per substrate atom.  Even if this information were available, there
are many other factors, discussed at the outset, which can contribute to the shift.
This is true especially for non-alkali adsorbates.  

Alkali adsorbates invariably lower the surface band, eventually dragging it into the bulk continuum, where it becomes a resonance; often the details are not reported (for example for Cs/Cu(111)\cite{Breitholtz:Cs_Cu111}); associated calculations are problematic due to the large unit cells needed for fractional coverage and ill-defined order.  There have been recent studies, using two-photon photoemission of all the alkalis on Cu(111)\cite{Zhao2008}
and on Ag(111)\cite{Wang2011}; they confirm the downward shift but offer little additional quantitative information on the coverage dependence of the shifts.  To address this shift quantitatively with DFT for Na/Cu(111) while keeping a manageable cell size, Caravati and Trioni\cite{NaCu111:Caratavi} used a jellium-like model with one-dimensional Chulkov potential\cite{Chulkov:MetSurf} and found $\eps_F -\eps(0)$ to be colinear for coverages $\Theta$ = 0, 0.06, and 0.14, of the form $-0.303 - 2.2\Theta$, with a slightly smaller negative slope when $\Theta$ = 0.25 was also included. For noble-gas adsorbates, the shifts are
large while the increase in effective mass is small, implying that more than
just field effects are involved.

In passing, we discuss a specific complication in deciphering the tabulated numbers:
the lateral dipolar interaction.  This effect can influence the field at other
sites.  Most significantly, as alkali atoms get close to each other,
the direct dipolar
repulsion becomes more important than the indirect, surface-state-mediated
interaction. This happens when the dipole is large enough to produce a
significant shift in the surface state.\cite{Fratesi:Na_on_Cu}

We note that the maximum downwards shifts possible for metallic surface states are approximatively set
by the $E=0$ value of the minimum surface-state energy
$\epsilon_F-\epsilon^{E=0}(0)$, for example, as measured in
Ref.~\onlinecite{Stark:2004} and reported in Table IV
(as the `clean' entry in the `E:substrate' rows). The maximum possible
upward shift in $\epsilon^{E\neq 0} (0)$ is instead approximatively
given by the difference between value of $\epsilon^{E=0}(0)$ and the energy $\varepsilon^{\rm bulk}(L)=900$meV
of the bulk state at the bottom of the Cu L-gap\cite{Knapp:PRB,Kevan:PRL}
(since the overlap in energies will convert the surface state to a surface-state resonance). The maximum upward
shifts of the surface state for Cu can therefore be estimated by
$\epsilon^{E=0}(0)-\varepsilon^{\rm bulk}(L)\approx$ 460meV.

When making precise use of the band shifts, one must take into account the temperature, as discussed in detail for the (111) faces of the three noble metals by Paniago et al.\cite{Paniago:Shokcley}  In particular, they show
$\epsilon_F-\epsilon^{E=0}(0)$ = $-$(75 $\pm$ 5) meV +(0.17 meV/K) $T$ for Ag(111).  There is a small increase also in $\hat{m}$, from 0.43 $\pm$ 0.04 at 65 K to 0.45 $\pm$ 0.04 at 294 K.    For Cu(111) the increase is comparable, with linear coefficient (0.18 $\pm$ 0.01) meV/K.

Overall, it is noteworthy that shifts far larger than those reported at the end of
Subsection \ref{moddisp} are seen. The adsorption-induced shifts are indeed larger than the shifts
of 50 meV which formed the abscissa limits in Fig.~\ref{fig:shift}.  Hence, the range of
field strengths investigated in our study are physically sensible.

\subsection{Decoupling method for adsorbed molecules}

Our study of response to an external field has been aided by the decoupling method described and tested in Section~\ref{Sec:decouple}.
This method can also be used to study shift in surface-state energy produced by an adsorbed molecule or atom.
This requires that the two KS states corresponding to the SL states do not hybridize with other bulk or molecular KS states. For systems with inversion symmetry in the plane (in terms of the basis vectors), like benzene on Cu(111),\cite{benzCu} we only need to decouple the states using an O(2) rotation.

As a test case, we consider benzene on Cu(111) in a 3{$\times$}3 periodic unit cell and a 6 layer slab. Using DFT with the vdW-DF2 functional to capture the non-local correlation essential to the binding of this system,\cite{vdWDF2} we determine a binding separation of $3.5 {\rm \AA}$, and a binding energy of 0.48 eV. Except for the use of vdW-DF2, the details of this calculation are the same as in our previously reported calculation with vdW-DF1.\cite{benzCu}
After decoupling the two KS states, we find a coupling of $\Omega=320$meV and a detuning of $W=16 {\rm meV}$. 
Since for the adsorbed-molecule case only one of the surfaces is perturbed, the
energy difference between the decoupled states $2W=32 {\rm meV}$ equals the shift in surface state energy.

In this calculation, the decoupling method enabled us to extract the shift in surface state energy in a traditional slab calculation using far fewer layers than what would be required for the KS to fully decouple because of the (weakly) broken symmetry. We have also used the method to study the surface shift induced by adsorbed anthraquinone chains on Cu(111) for varying coverages.\cite{Wyrick:Do2d}

\section{Conclusions}

Using DFT calculations, we have shown that an electric field perpendicular to a
metal surface with a metallic Shockley surface state linearly shifts the bottom
of this state (relative to the Fermi energy) for physically plausible field
strengths.  We have computed the value of the linear proportionality constant,
as well as that associated with the field-induced change in the curvature of the
dispersion relation, that is, in the effective mass.

The decoupling method presented here should be useful when studying the response
in the surface-state dispersion to a perturbation that does not destroy the
surface-state character. It is, for example, relevant for a study
of the surface response arising from organic overlayers weakly coupled to the surface or dilute overlayers of
chemisorbed atoms.

More generally, the MaxLoc analysis could be useful for decoupling
states which arise at different spatial locations and which hybridize
under the asumption that an infinite time is available to create the coupling.
The characterization of the coupling in such systems can be used to
calculate tunneling rates and oscillation frequencies.

It is hard to overemphasize the significance of acquiring the ability to control
the Fermi wavelength in order to manipulate and engineer surface structures
determined by interactions mediated by surface states.  With a strong enough
field, one could in principle manipulate channels in a manner reminiscent of
Repp's resonator\cite{Negulyaev:DirectEvidence} and so dynamically direct an
atom flow on surface.

\section{Acknowledgement}
The authors thank the Swedish National Infrastructure for Computing (SNIC) for
access and for KB's  participation in the graduate school NGSSC. The work at
Chalmers was supported by the Swedish research Council (Vetenskapsr\aa det VR)
under  621-2008-4346 and by VINNOVA.  Work at U. of Maryland was supported by
NSF Grant CHE 07-50334 and by NSF-MRSEC Grant DMR 05-20471 and ancillary support
from the Center for Nanophysics and Advanced Materials (CNAM) and a DOE CMCSN
grant.

\bibliographystyle{apsrev}	
\bibliography{extE2_kb1}		%

\begin{thebibliography}{128}
\bibitem[{Com({\natexlab{a}})}]{Comment_density}
\bibinfo{note}{The corresponding cutoff for the density was set to 500 eV}.
\bibitem[{Com({\natexlab{b}})}]{Comment_rel}
\bibinfo{note}{In preliminary low-accurcy calculations, the max force was 0.055
  for the smallest electrical field, and 0.057 for the largest. For a high
  accurcay calculations at zero electical field, the max force was 0.042,
  smaller than the standard limit of 0.05. Thus we do not include relaxations
  as they are prone to introduce small kinks in the curves.}
\end{thebibliography}

\end{document}